\RequirePackage{fix-cm}

\documentclass[twocolumn]{svjour3}          
\usepackage[utf8]{inputenc}
\usepackage{longtable,multirow,supertabular,afterpage}
\usepackage{amssymb,amsbsy,textcomp,marvosym,caption,threeparttable}
\usepackage{mathptmx, amsmath, amsfonts}
\usepackage{bbm} 
\usepackage{placeins}  
\usepackage{nicefrac,xfrac}
\usepackage{eurosym,fancyhdr,CJK,multicol,graphics,indentfirst,color,bm,upgreek,booktabs,graphicx,subfigure}
\usepackage[mathcal]{euscript}
\usepackage{setspace}
\usepackage{float}
\usepackage{hyperref}

\usepackage{array}
\usepackage{enumitem}

\usepackage[backend=bibtex,style = nature,sorting=none]{biblatex}
\newcommand\mybullet{$\bullet$~}

\usepackage{makecell}

\newcolumntype{P}[1]{>{\centering\arraybackslash}p{#1}} 
\newcolumntype{L}[1]{>{\raggedright\arraybackslash}p{#1}} 
\usepackage{hyperref}
\usepackage{graphicx}
\usepackage{titlesec}

\begin{document}
\title{Role and Integration of Image Processing Systems in Maritime Target Tracking}

\author{Yassir Zardoua$^\star$\and Bilal Sebbar\and Moussab Chbeine\and Abdelali Astito\and Mohammed Boulaala}

\institute{Yassir Zardoua \textbf{(Corresponding author)}\at
	\email{yassirzardoua@gmail.com}    \\       
	$^\star$ Smart Systems \& Emerging Technologies, FSTT, Abdelmalek-Essaadi University, Tetouan, Morocco\\
}

\maketitle

\begin{abstract}	
In recent years, maritime traffic has increased, especially in seaborne trade. To ensure safety, security, and environmental protection, various systems have been deployed, often combining data for improved effectiveness. One key application of this combined data is tracking targets at sea, where the Automatic Identification System (AIS) and X-band marine radar are crucial. Recently, there has been growing interest in using visual data from cameras to enhance tracking. This has led to the development of several tracking algorithms based on image processing. While much of the existing literature addresses data fusion, there hasn't been much focus on why integrating image processing systems is important given the existence of the other systems. In our paper, we aim to analyze these surveillance systems and highlight the reasons for integrating image processing systems. Our main goal is to show how this integration can improve maritime security, offering practical insights into enhancing safety and protection at sea.
\end{abstract}

\section{Introduction}
\begin{figure*}[h]
	\centering
	\includegraphics[width= 0.7\linewidth]{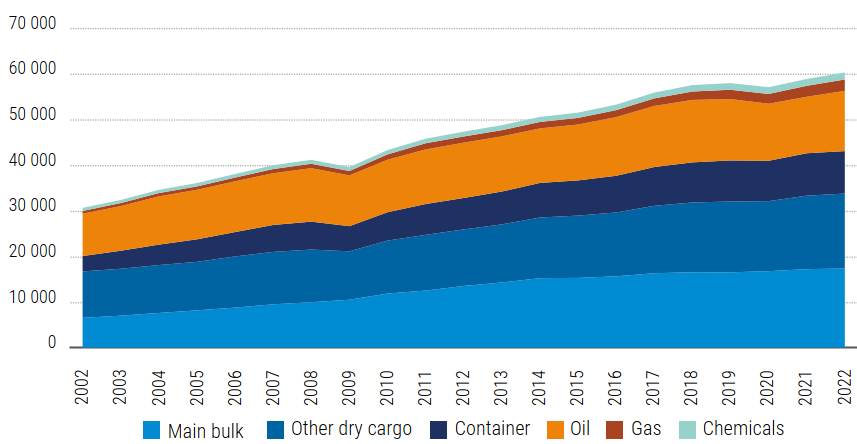}
	\caption{Global maritime commerce: Cargo ton-miles in the billions, spanning 2002 to 2022. Source:~\cite{trade_statistics}}
	\label{fig2}
\end{figure*}

Maritime movement is concentrated, particularly in straits and certain coastal regions. Statistical data~\cite{book1} confirms a continuous rise in maritime traffic, particularly in the context of trade activities (see Figure~\ref{fig2}). Consequently, the occurrence of various threats is anticipated~\cite{androjna2022cyber, liss2022maritime}. To prevent and mitigate the impact of maritime threats, well-defined missions must be carried out, primarily stemming from conventions and regulations established by the IMO (International Maritime Organization)~\cite{imo}. Various centers, such as VTS (Vessel Traffic Service), FMC (Fishery Monitoring Center), MMC (Mission Control Center), are established by contracting governments to contribute to maritime surveillance at both national and international levels~\cite{payne2023marine}. In maritime surveillance, threats are primarily averted through the detection of anomalies. In the context of surveillance, an anomaly is an abnormal behavior, which can be detected through the  identification unusual patterns in collected data~\cite{anomaly_review, gamage2023comprehensive} or uncovering behaviors that are not typically observed.

The detection of the aforementioned anomalies can be accomplished through various methods, one of which involves the analysis of data collected from surveillance systems. These systems yield multiple categories of data. In this paper, we specifically focus on the use of tracking data for maritime surveillance. Such data is collected from various surveillance systems, with our primary focus being on AIS (Automatic Identification System), X-band marine Radar systems, and image processing systems. Optical satellite systems (e.g., Quick Bird and SPOT) and Radar satellite systems (e.g., SAR) are excluded from our study due to their extended temporal resolution, rendering the collection of tracking data impractical

The remainder of the paper is structured as follows: In Section~\ref{section2}, we clarify the significance of tracking data and its importance in the detection of maritime threats. Section~\ref{section3} addresses the limitations of AIS and X-band marine radar in collecting tracking data and underscores the role of image processing systems in mitigating these limitations. We conclude in Section~\ref{conclusion} by emphasizing the enhancements in maritime security that arise from the integration of image processing systems.

\section{Role of Tracking Data in Anomaly Detection}
\label{section2}
This section defines the tracking data and explain its relevance to the prevention of maritime threats through anomaly detection. The reader can find summarized highlights of this section in Table~\ref{tab_summ_tracking_data}.
\begin{table*}[!h] 
	\centering
	\caption{\label{tab_summ_tracking_data}\footnotesize{A summary of the key concepts and components related to the role of tracking data in anomaly detection}}\vspace{-2mm}
	\setlength{\extrarowheight}{5pt}
	{\footnotesize
		\begin{tabular}{
				|L{0.2\linewidth}||
				L{0.65\linewidth}|}
			\hline
			Aspect&Description\\
			\hline
			\hline
			Anomalies detection&Performed through analysis of ships' tracks exhibiting unusual patterns or breaching navigation rules.
			\\
			\hline
			Ship tracking steps& Target detection, recognition, and identification.
			\\
			\hline
			Target detection& Involves indicating the presence of an object and estimating its location.
			\\
			\hline
			Target recognition& Recognition classifies the detected target as a ship and identifies its type (e.g., fishing boat, cruise ship).
			\\
			\hline
			Target identification&Involves labeling ships with unique identifiers and updating their locations.
			\\
			\hline
			Ideal identifier& Ideally, ships are identified by their IMO number, serving as a permanent unique reference for a ship. However, when IMO numbers are unavailable, system-specific identifiers are assigned.
			\\
			\hline
			COLREGs convention	&The COLREGs convention provides detailed safety rules for maritime traffic, emphasizing the role of tracking data.
			\\
			\hline
			Effective anomaly detection	&Combining predefined patterns and safety rules, including those from COLREGs, provides effective anomaly detection.
			\\
			\hline	
		\end{tabular}
	}
\end{table*}

Anomalies can be detected through the analysis of various types of data and events, with ship tracking data being particularly relevant in this context. Tracking a ship typically involves three major steps: detection, recognition, and identification~\cite{huang2021identity, liu2019ship, park2022object}. Target detection entails indicating the presence of an object and estimating its location. Subsequently, it's essential to recognize it as a marine vessel. Target recognition aims to classify the detected target as a ship and then identify its type (e.g., fishing boat, bulk ship, cruise ship). The identification step is useful when multiple ships can be detected simultaneously, tracking a specific target requires labeling each ship with a unique identifier and update its location with a suitable frequency. Ideally, ships are identified by their IMO number, which serves as a unique permanent reference for a ship. However, in many cases, the IMO number is unknown to the tracking system, or the vessel may not even have an IMO number. In such instances, a unique identifier recognized by the surveillance system is used. With that in mind, tracking data includes the vessel's type or class, an identifier to distinguish it from other vessels of the same type, and a track, which comprises a history of all its previous positions.

Tracking data is utilized for comparison with predefined patterns or rules~\cite{gamage2023comprehensive}, both of which are effective in identifying anomalies. The first method is employed due to the fact that a ship's track is characterized by a specific set of patterns, primarily determined by its type or the nature of its activities~\cite{anomaly_detection}. For instance, vessels engaged in international cargo transportation, such as bulk carriers, tend to follow the most efficient route from departure to destination~\cite{zhang2019data, ozturk2022review}. Regular detections of a recognized ship of a particular type are accumulated to create a track. Multiple tracks of ships engaged in similar activities are aggregated to construct a model of normalcy, representing typical tracks. Consequently, a ship that deviates from this normal track is considered a potential threat~\cite{nguyen2021geotracknet, yan2019study}.

The second method is comparatively quicker and more straightforward for anomaly detection. It involves the establishment of safety rules~\cite{nemeth2019private}, and any breach of these rules is considered an anomaly. An instance of this is the 'zone entry anomaly', which involves verifying whether a ship of a specific type has entered a designated zone~\cite{anomaly_detection}. For instance, remote coastal areas are typically frequented by medium and large-sized vessels. The detection of a small boat in such an area is atypical, thus constituting a 'zone entry anomaly,' which may be indicative of activities like illegal drug trafficking or immigration.

Safety rules can be quite detailed, as exemplified by those outlined in the COLREGs (COLlision REGulations) convention~\cite{COLREGs}. This convention is designed to facilitate efficient maritime traffic, diminish the likelihood of collisions, and prevent unauthorized boardings. The rules prescribed by COLREGs predominantly pertain to safe speeds, permissible maneuvering, right of way, overtaking procedures, and similar considerations. Additional rules encompass the assessment of collision or boarding risks based on the speed and heading of nearby vessels, along with appropriate actions to be taken in the event of a collision alert. The nature of these regulations underscores the critical role of tracking data in their enforcement.

In essence, the analysis of tracking data plays a crucial role in maritime surveillance, facilitating the detection of anomalies that could present security risks. The integration of predetermined patterns and safety rules, influenced by conventions like COLREGs, offers effective approaches to detect such anomalies.

\section{Collection of Tracking Data: Comparison of the Main Systems}
\label{section3}
In this section, we will compare the usage of AIS, radars, and image processing systems in the collection of tracking data. Within this comparison, we will highlight the limitations of AIS systems and demonstrate how radars can, to some extent, mitigate these limitations. Image processing systems are introduced as a complement to AIS and radars, and we will eventually illustrate the resulting improvements derived from using image processing systems. The reader can refer to Table~\ref{tab_summ_analysis} for quick highlights of this section.

\begin{table*}[!h] 
	\centering
	\caption{\label{tab_summ_analysis}\footnotesize{A summary of the effectiveness of the three systems in the target tracking task}}\vspace{-2mm}
	\setlength{\extrarowheight}{5pt}
	{\footnotesize
		\begin{tabular}{
				|L{0.1\linewidth}||L{0.22\linewidth}|L{0.24\linewidth}|L{0.24\linewidth}|}
			\hline
			System&Description&Advantages&Inconveniences\\
			\hline
			\hline
			AIS (Automatic Identification System)&
			Data collection via VHF signals from special devices
			&
			\mybullet Wide coverage area\newline
			\mybullet Real-time data \newline
			\mybullet Specific vessel mandates
			&
			\mybullet Not all vessels required to use AIS \newline
			\mybullet Potential AIS switch-off \newline
			\mybullet Limited to certain vessel classes \newline
			\mybullet Update rates may decrease \newline
			\mybullet Occasional positional errors\\
			\hline
			X-Band Marine Radar&
			Emit signals in the 8.0 to 12.0 GHz range and measure returns in terms of signal strength and frequency shifts&
			\mybullet Real-time data \newline
			\mybullet No special device on the target is required \newline
			\mybullet High update rate
			&
			\mybullet Limited small target detection \newline
			\mybullet Limited target recognition\\
			\hline
			Image processing systems	&
			Visual data collection via cameras and image processing with computers&
			\mybullet Detection of small targets \newline
			\mybullet More relevant data for classification and recognition\newline 
			\mybullet Optics allow long-range detection
			&
			\mybullet Limited to short coverage areas (compared to the AIS)\newline
			\mybullet Data processing overhead\newline
			\mybullet Effective recognition requires advanced algorithms 
			\\
			\hline
		\end{tabular}
	}
\end{table*}

\subsection{AIS (Automatic Identification System)}
AIS is the primary system employed in maritime surveillance~\cite{international2015revised, liu2019intelligent}, providing tracking data for ships and other information categorized into four main groups:

\begin{enumerate}
	\item Static Information: This category includes details such as the vessel's class, name, flag, image, IMO (International Maritime Organization) and MMSI (Maritime Mobile Service Identity) number, GT (Gross Tonnage), and dimensions.
	\item Dynamic Information: Dynamic information encompasses position, speed, acceleration, and track data.
	\item Voyage-Related Information: This category contains information related to the type of cargo, number of passengers, destination, ETA (Estimated Time of Arrival), and route plan.
	\item Short Safety-Related Messages: This group includes critical safety messages, such as information about tides, weather conditions in specific areas, and warnings related to events like suspected cy or terrorist activities.
	
\end{enumerate}

\begin{figure*}[h]
	\centering
	\includegraphics[width= 1\linewidth]{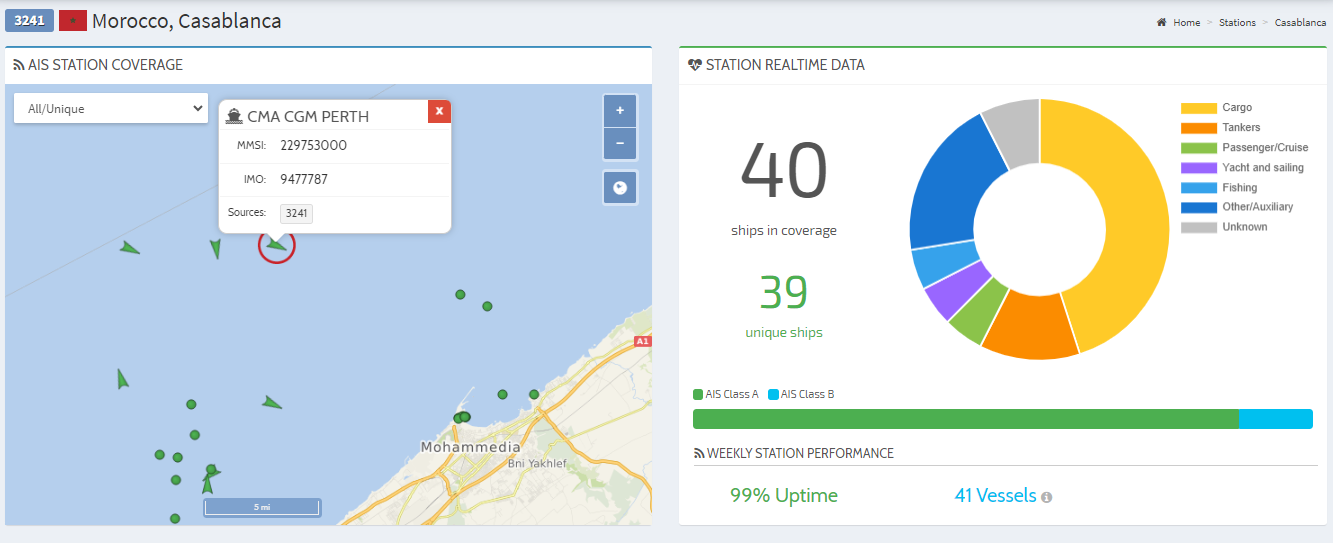}
	\caption{The AIS data from the Moroccan station Casablanca:. Source: \url{https://www.aishub.net/stations/3241}}
	\label{fig_ais_online}
\end{figure*}

The AIS data of ships worldwide is available online as shown in Figure~\ref{fig_ais_online}. This data is sourced from various providers, including weather stations and ship-based sensors like GPS, or it may be manually logged by the ship's officers. The sharing and access to AIS information are facilitated through VHF coast stations. There are several limitations associated with AIS in collecting tracking data. According to SOLAS (Safety of Life at Sea) regulations~\cite{SOLAS}, not all ships are required to transmit AIS signals. AIS is mandated solely for passenger ships, vessels over 300 GT on international voyages, and vessels over 500 GT on non-international voyages, with naval vessels being excluded. Furthermore, there is no guarantee that vessels mandated to transmit AIS signals will consistently comply, as the AIS device can be switched off. In addition, AIS information may suffer from slow update rates and occasional positional errors. These limitations give rise to three significant inconveniences in target tracking: notable inaccuracies in detecting fast-moving vessels, the non-detection of non-mandated AIS-equipped boats, particularly small boats, and the inability to detect ships that have deactivated their AIS. The subsequent Section~\ref{radar}, will delve into how these issues can be mitigated using radar systems.

\subsection{X-band Marine RADARs}
\label{radar}

Marine radar is an instrument operating within X-band frequencies (8.0 to 12.0 GHz). It employs a rotating flat antenna that continuously scans a narrow beam of microwaves across all horizontal directions. The same antenna detects reflected waves, facilitating the identification of surrounding obstacles and marine vessels, displayed on a screen. Unlike AIS, radar systems do not require targets to have special devices, enabling the detection of ships not transmitting the AIS signal. Radar systems provide a high detection update rate, with updates occurring every 5 seconds, compared to AIS, which may have updates as infrequent as 120 seconds~\cite{sagild2021track}. This capability effectively addresses the challenge of tracking fast-moving vessels.

Despite the advantages of RADAR systems in comparison to AIS, they are not without their drawbacks. Radar systems have been criticized for their limitations in detecting small targets, and they have been implicated in several accidents involving small boats~\cite{accident_report}. An analysis of Search and Rescue (SAR) actions conducted in the Adriatic Sea suggests that small boat accidents represent a significant portion of the total number of incidents~\cite{small_craft_role}. A recent review~\cite{zainuddin2019maritime} highlights that small vessel detection remains an ongoing challenge for marine radars and an open issue that is continuously being explored by scholars. Another limitation of radar systems is their limited ability to classify the detected target, which is critical information, as explained in Section~\ref{section2}. While this limitation can be mitigated if the target is transmitting the AIS signal, as it contains information about the type and activity of the vessel, the issue persists because this information can be falsified or the vessel may not be equipped with an AIS device.

In the remainder of this section, we delve into the reasons behind the limitations of marine radars in detecting small targets. Radar Cross Section (RCS) serves as a measure of the electromagnetic signal reflectivity of an object, predominantly dependent on the object's size, material, and shape~\cite{charris2012analysis}. Objects with a low RCS exhibit weak signal reflections. On the other hand, sea clutter is any undesired signal reflection stemming from the nature of the sea. Capillary waves and gravity waves are primarily induced by winds and are recognized as the primary sources of sea clutter for X-band radars~\cite{raynal2010doppler}.

\begin{figure*}[h]
	\centering
	\includegraphics[width= 0.7\linewidth]{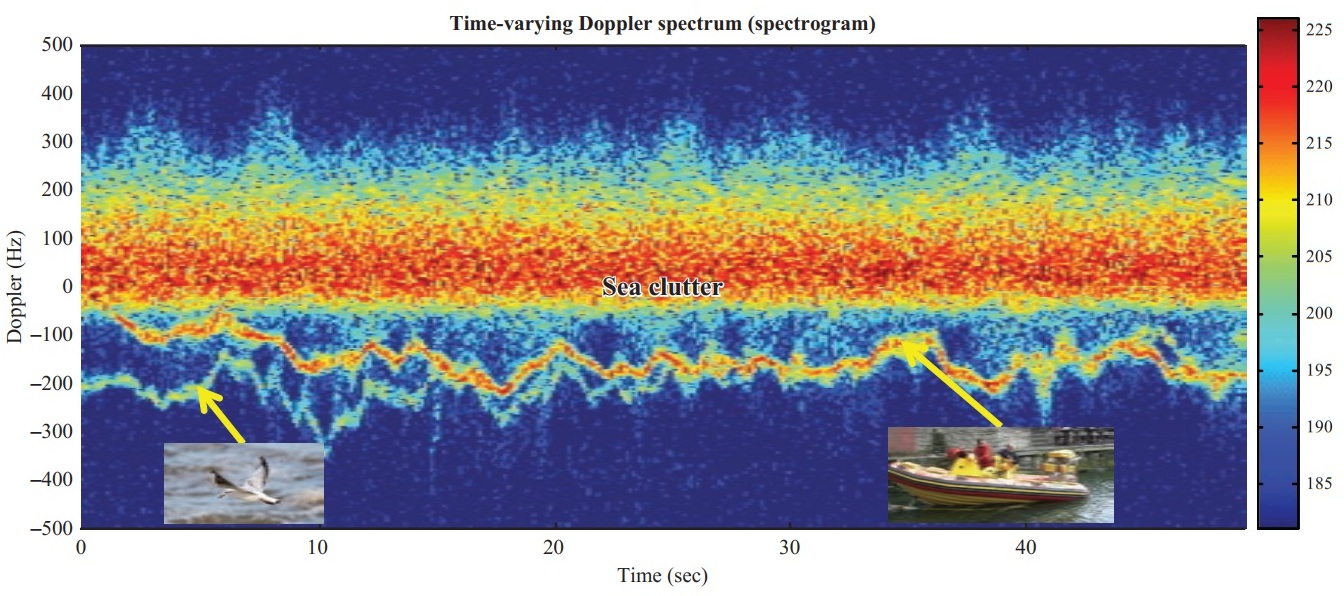}
	\caption{The time-varying Doppler frequency shift of the small boat, sea clutter and flying birds~\cite{chen2014radar}}
	\label{fig3}
\end{figure*}

The detection of small targets is challenging due to the low Signal to Noise Ratio (SNR) attributed to sea clutter and the low Radar Cross Section (RCS) values associated with small targets. Radar techniques utilizing the Doppler effect have proven to be effective in small target detection amidst sea clutter~\cite{herselman2008analysis}. The Doppler effect occurs when there is a change in the distance between the radar transmitter and the target. This change results in a shift in the received frequencies, known as the Doppler frequency shift, which is determined by the radial velocity of the target~\cite{chen2019micro}. When the Doppler frequency shifts of sea clutter and small targets do not overlap, small object detection becomes feasible, as illustrated in Figure~\ref{fig3}.

In some cases, the Doppler shift of certain targets, such as the RIB (Rigid Inflatable Boat) and the seagull after the time instant $t = 30 sec$ (as shown in Figure~\ref{fig3}), may exhibit overlapping bands, thereby complicating the differentiation between targets. This intersection can also occur between sea clutter and targets of interest~\cite{raynal2010doppler}, potentially obscuring the Doppler shift associated with small vessels~\cite{chen2014radar}, rendering their detection impractical.

\subsection{Image Processing Systems}
\subsubsection{Cameras as a Complimenting Sensor}
We have observed that the combination of AIS with radar systems has limitations, particularly in collecting the tracking data of ships that do not have AIS equipment, as well as in detecting small targets. Several studies suggest that cameras are promising candidates for complementing existing surveillance systems~\cite{small_craft_role,ponsford2001integrated,almeida2009radar, zardoua2021survey}.

Recent advancements in imaging technology have made cameras strong contenders for integration with other technologies. These developments encompass high-resolution imaging, the availability of flexible lenses for adjusting the field of view, and the capability to capture visual data across various light frequencies, including the infrared spectrum, which is particularly valuable for night vision.

Data obtained from vision sensors are also well-suited for automatic processing. This capability proves invaluable in addressing challenges associated with (i) human errors resulting from fatigue and information overload, and (ii) the resource requirements, including the number of watch-standers necessary to monitor multiple Closed Circuit Television (CCTV) screens, as well as their training.

In the context of target tracking, image processing systems offer two distinct advantages when compared to the combination of AIS and radar systems. The first advantage is the enhanced probability of detecting small targets, as they are readily visible in images. The second advantage is the capability to recognize vessel types. Section~\ref{3.3.2} provides an illustration of how these advantages can contribute to enhancing the security of the maritime environment.

\subsubsection{Deployments of Camera-Based Surveillance Systems and the Resulted Security Improvements}
\label{3.3.2}
Considering the advantages presented by vision sensors in terms of target tracking, several systems that integrate cameras as supplementary sensors have been developed to enhance the detection of maritime threats~\cite{seecoast,maaw,argos,aiv3s,asv}. To demonstrate the security improvements achieved, it's essential to examine various deployments of camera-based surveillance systems. We will focus on three primary deployment types: ground-based, buoy-based, and ship-based video surveillance.

\paragraph{Ground-Based Video Surveillance}
In Section~\ref{section2}, we explained that the detection of anomalies with tracking data primarily involves comparing them with a set of patterns and rules. Although automatic video surveillance may currently be impractical for collecting data across a wide coverage area, such as tracking a bulk carrier on an international voyage, it can be effectively deployed for data collection within shorter coverage areas (e.g., 5km to 10km)~\cite{auslander2011comparative}, such as ports, harbors, and rivers. This localized deployment allows for the collection of tracks of small targets and the recognition of vessel types. Such capabilities significantly enhance the ability to predict threats, as this task necessitates access to track data and the knowledge of marine vehicle types.

\paragraph{Buoys-based video surveillance}
Buoys-based video surveillance entails the establishment of a network of buoys, each equipped with a camera, a processor for image processing tasks, and a bi-directional communication unit for transmitting the collected information to surveillance centers~\cite{fefilatyev2012detection,zhang2017ship}. With the implementation of appropriate processing algorithms, these systems can be effectively deployed in open-ocean environments to detect and identify small boats, which are often associated with illegal immigration and drug trafficking. Another notable enhancement involves the prevention of poaching, particularly in cases where the VMS (Vessel Monitoring System) device of a ship is turned off. This can be achieved through the recognition of the vessel as a fishing ship operating within a restricted fishing area.

\paragraph{Ship-based video surveillance}
As small boats typically do not carry AIS and are less likely to be detected by marine radars, cameras can serve as a valuable complement to a ship's navigation equipment. This integration proves highly beneficial in mitigating the risk of collisions and preventing maritime threats in open ocean environments, particularly acts of piracy and terrorist attacks, which, based on several incidents, are frequently carried out using small boats~\cite{hill2009maritime, jin2019marine}. Additional enhancements encompass the ability to conduct search and rescue operations for individuals in distress, especially when utilizing cameras operating in the infrared spectrum. Infrared cameras offer excellent human body contrast~\cite{asv}, facilitating the easy detection of individuals in need of assistance.

\section{Summary and discussion}
\label{conclusion}
In conclusion, we have analyzed the role of tracking data in maritime anomaly detection, comparing the capabilities and limitations of AIS, X-band marine radars, and image processing systems. We have identified that while AIS is essential for tracking data, it has shortcomings in tracking non-mandated vessels and small boats. X-band marine radars offer an alternative but face issues with small target detection and classification. 

Image processing systems emerge as valuable complements to AIS and radar. Recent technological advancements, including high-resolution imaging and infrared spectrum capabilities, position cameras as effective tools for detecting small targets and recognizing vessel types. These systems play a pivotal role in enhancing maritime security, addressing the limitations presented by AIS and radar systems. 

The integration of image processing systems contributes significantly to the detection of maritime threats, including small vessels and potential security risks such as piracy and illegal activities. By providing high-resolution images and enabling automatic data processing, these systems offer a comprehensive solution to strengthen maritime security.

\printbibliography
\end{document}